\documentstyle[11pt,aas2pp4,flushrt]{article}

\lefthead{Nakariakov, Mendoza-Brice\~no \& Ib\'a\~nez}
\righthead{Shock Wave Formation}

\begin{document}

\title{Magneto-Acoustic Waves of Small Amplitude\\
in Optically Thin Quasi-Isentropic Plasmas}
\author{Valery M. Nakariakov}
\affil{School of Mathematical and Computational Sciences, University
of St~Andrews, St~Andrews, Fife KY16 9SS, Scotland, UK}

\and

\author{ C\'{e}sar A. Mendoza-Brice\~{n}o, Miguel H.
Ib\'{a}\~{n}ez S.}
\affil{Centro de
Astrof\'{\i}sica Te\'{o}rica, Universidad de los Andes, Apartado de Correos
26, IPOSTEL La Hechicera, M\'{e}rida, Venezuela}  

\begin{abstract}
The evolution of quasi-isentropic magnetohydrodynamic waves of small but
finite amplitude in an optically thin plasma is analyzed. The plasma is assumed
to be initially homogeneous, in thermal equilibrium and with a straight
and homogeneous magnetic field frozen in. Depending on the particular form of
the heating/cooling function, the plasma may act as a dissipative or active
medium for magnetoacoustic waves, while Alfv\'{e}n waves are not directly
affected. An evolutionary equation for fast and slow magnetoacoustic waves in
the single wave limit, has been derived and solved, allowing us to analyse
the wave modification by competition of weakly nonlinear and quasi-isentropic
effects.  It was shown that the sign of the quasi-isentropic
term determines the scenario of the evolution, either dissipative or active. In
the dissipative case, when the plasma is first order isentropically stable the
magnetoacoustic waves are damped and the time for shock wave formation is
delayed.  However, in the active case
when the plasma is isentropically overstable, the wave amplitude grows, the
strength of the shock increases and the breaking time decreases. The magnitude
of the above effects depends upon the angle between the wave vector and the
magnetic field. For  hot ($T>10^{4}$ $K$) atomic plasmas with solar
abundances either in the interstellar medium or in the solar atmosphere, as
well as for the cold ($T<10^{3}$ $K$) ISM molecular gas, the  range
of temperature where  the plasma is  isentropically unstable and the
corresponding time and length-scale for wave breaking have been found.
\end{abstract}


\section{Introduction}

The study of wave propagation and the stability of optically thin plasmas is
of crucial importance for understanding the origin and evolution of
inhomogeneities observed at very different length scales and time scales in
astrophysical plasmas. Travelling acoustic waves in the plasma can be
unstable due to peculiarities of the processes of heating and cooling of the
medium. The possible amplification of the waves is connected with additional
energy liberation in the plasma in the compression phase or, in other words,
with nonadiabaticity of the plasma. In different astrophysical applications,
the nonadiabaticity of the plasma can be mathematically modelled by an
introduction of an additional term in the adiabatic equation. This term,
usually called a heating/cooling function, depends on parameters of the
plasma under consideration. The specific expression for this function varies
in different physical situation (Rosner, et. al., 1978; Vesecky et. al.,
1979; Dahlburg and Mariska, 1988), but fortunately some general properties
of this function define main features of dynamics of perturbations in
optically thin plasmas (either amplification or decay).

The above problems, taking into account many different physical effects have
been extensively studied as far as the linear approximation is concerned
(Parker, 1953; Kruskal and Schwarzschild, 1966; Weymann, 1960; Field, 1965;
Priest, 1982; Ib\'a\~nez and Parravano, 1983; Ib\'a\~nez, 1985; Iba\'a\~nez
and Escalona, 1993; Bodo et al.,1985; Balbus and Soker, 1989 and 
references therein). The linear
approach corresponds to the case of waves with zero amplitude and,
consequently, can describe only an initial stage of the instabilities
associated with nonadiabatic effects in the plasma. Generally, the nonlinear
regime is not far well understood and most of the progress has been done by
numerical simulations of very particular structures (Dahlburg et al., 1987;
Karpen et al., 1989; Reale et al., 1994).

In addition to condensations that can be formed in hot plasmas by thermal
instability in a nonadiabatic plasma, providing that the instability does
not saturate in the nonlinear regime, the steeping of travelling waves may
also originate inhomo\-geneities (Oppenheimer, 1977; Krasnobaev, 1975;
Krasnobaev and Tarev, 1987 (Reference KT herein after); Tarev, 1993). This
problem has been worked out in KT and Tarev (1993) references neglecting the
presence of magnetic fields. However, in many astrophysical situations the
magnetic fields become relevant, therefore, it is worthy to analyse the
propagation of nonlinear magnetoacoustic waves in plasmas where nonadiabatic
effects also go into play. This problem beyond its heuristic interest, also
can be the clue for understanding particular observed features in magnetised
plasmas, for instance, the intensity oscillations observed in the network
bright points in the quiet Sun (Kalkofen, 1997). Also, in the presence of
the magnetic field, the magnetoacoustic waves perturb the absolute value and
components of the field, which gives an additional possibility for the
registration of the discussed phenomena in astrophysical plasmas through
observation of variation of gyro-emission, Zeeman effect etc.

The aim of the present paper is the general study of weakly nonlinear
magnetoacoustic and Alfv\'{e}n wave dynamics in presence of nonadiabatic
effects in an optically thin plasma. We consider a homogeneous plasma 
penetrated by the straight and homogeneous magnetic
field. Effects of self-gravitation, ionization and steady flows are
neglected. The physical state is determined by time and spatial independent
variables. Additionally, the heating/cooling function (per unit volume and
time) is assumed to be dependent of pressure and density, i.e. $Q(p,\rho )$.
Only quadratic nonlinear effects will be taken into account. Assumption that 
nonlinear and nonadiabatic effects are weak, but not necessarily of the same
order, allows us to apply the method of slowly varying amplitudes and  derive
an evolutionary equation for fast and slow magnetoacoustic waves. Note that,
despite the nonlinearity considered in the paper is weak, it leads to the
significant modification of the wave, up to formation of shocks. In the
derivation of the evolutionary equation, a single wave approximation is
applied. This means that interaction of waves of different types, as well as
interaction of forward and backward waves, is
neglected.  The nonlinear interaction of different modes, for example,
three-wave interaction, occurs when all interacting modes are existing in the
system (and are, actually, over some certain amplitude threshold prescribed by
dissipation). Also, efficiency of the interaction depends  strongly on
coherence of the interacting waves. Both these conditions  are assumed to
be not fulfilled: we consider a wave
of one certain type only, while other modes are not excited. Also, our
attention is restricted to evolution of wave pulses with a wide spectrum, and
nonlinear resonant interactions are suppressed. Thus, the
dominating nonlinear process is generation of higher harmonics, leading to wave
steepening and shock formation.  Moreover, this process is more pronounced due
to the absence of linear high frequency dispersion. Nonlinear dispersion
preventing the second harmonics generation, may appear in the next order of
nonlinearity and  is out of scopes of the paper.

Since the
thermally unstable plasma is an {\it active} medium for the magnetoacoustic
waves, the solution of the evolutionary equation shows the amplification of
the initial magnetoacoustic perturbation and speeding up of shock formation.
The breaking time depends on the angle of the wave propagation with respect to
the magnetic field and is different for slow and fast waves. The above results
are quite general and potentially can be applied to different
astrophysical as well as laboratory plasmas. However, in this paper, the
results obtained will be addressed to study a hot ($T>10^{4}$ $K{\cal
)}$ atomic plasma (Solar atmosphere or hot Interstellar Medium) and a cold
($T<10^{3}$ $K{\cal )}$ molecular gas (say for instance, the cold ISM gas).

\section{Governing equations}

In the model considered, dynamics of the plasma is described by the set of
MHD equations,

\begin{equation}
\frac{\partial {\bf B}}{\partial t} = {\rm curl}\ {\bf V} \times {\bf B},
\end{equation}

\begin{equation}
\rho \frac{d {\bf V}}{d t} = -\nabla p - \frac{1}{4\pi} {\bf B}\times {\rm %
curl} \ {\bf B},
\end{equation}

\begin{equation}
\frac{\partial \rho}{\partial t} + {\rm div}\ \rho {\bf V} = 0,
\end{equation}

\begin{equation}
\frac{d p}{d t} - \frac{\gamma p}{\rho} \frac{d \rho}{d t} = (\gamma-1)\
Q(p,\rho),
\end{equation}

\begin{equation}
{\rm div} {\bf B} = 0,
\end{equation}
where $Q(p, \rho)$ is a known function describing effects of the
nonadiabaticity and all other notations are standard.

The stationary magnetic field is in the $xz-$plane, ${\bf B_0} = B_0 {\rm sin%
} \alpha\ {\bf x_0} + B_0 {\rm cos} \alpha {\bf z_0}$, where $B_0$ is the
absolute value of the magnetic field, $\alpha$ is the angle between the
magnetic field and $z-$axis, ${\bf x_0}$ and ${\bf z_0}$ are the unit
vectors. We consider dynamics of waves propagating along the $z-$axis.
Dependences upon $x$ and $y$ are ignored ($\partial/\partial x =
\partial/\partial y = 0$).

Projecting Eqs. (1)---(4) on the axes, we have 
\begin{equation}
\frac{\partial B_x}{\partial t} = -\frac{\partial}{\partial z} ( V_z B_x -
V_x B_z),  \label{Bx}
\end{equation}

\begin{equation}
\frac{\partial B_y}{\partial t} = \frac{\partial}{\partial z} ( V_y B_z -
V_z B_y),
\end{equation}

\begin{equation}
\frac{\partial B_z}{\partial t} = 0,  \label{omit}
\end{equation}

\begin{equation}
\rho \frac{d\ V_x}{d\ t} = \frac{1}{4\pi}B_z\frac{\partial B_x}{\partial z},
\end{equation}

\begin{equation}
\rho \frac{d\ V_y}{d\ t} = \frac{1}{4\pi}B_z\frac{\partial B_y}{\partial z},
\end{equation}

\begin{equation}
\rho \frac{d\ V_z}{d\ t} = - \frac{\partial p}{\partial z} - \frac{1}{4\pi}%
\left(B_x\frac{\partial B_x}{\partial z} + B_y \frac{\partial B_y}{\partial z%
}\right),
\end{equation}

\begin{equation}
\frac{\partial \rho}{\partial t} + \frac{\partial}{\partial z} \left(\rho\
V_z \right),
\end{equation}

\begin{equation}
\frac{\partial p}{\partial t} + V_z \frac{\partial p}{\partial z} - \frac{%
\gamma p}{\rho}\left( \frac{\partial \rho}{\partial t} + V_z \frac{\partial
\rho}{\partial z}\right) = (\gamma-1)Q(p,\rho).  \label{p}
\end{equation}

We rewrite equations (\ref{Bx})-(\ref{p}) gathering linear and nonlinear
terms on left and right handside, respectively, 
\begin{equation}
\frac{\partial B_x}{\partial t}+\frac \partial {\partial z}\left(
B_{0x}V_z-B_{0z}V_x\right) =N_1,  \label{ms1}
\end{equation}

\begin{equation}
\frac{\partial B_y}{\partial t} - \frac{\partial}{\partial z} \left( B_{0z}
V_y\right) = N_2,  \label{a1}
\end{equation}

\begin{equation}
\rho_0 \frac{\partial V_x}{\partial t} - \frac{B_{0z}}{4\pi}\frac{\partial
B_x}{\partial z} = N_4,  \label{ms2}
\end{equation}

\begin{equation}
\rho_0 \frac{\partial V_y}{\partial t} - \frac{B_{0z}}{4\pi}\frac{\partial
B_y}{\partial z} = N_5,  \label{a2}
\end{equation}

\begin{equation}
\rho_0 \frac{\partial V_z}{\partial t} + \frac{\partial p}{\partial z} + 
\frac{B_{0x}}{4\pi}\frac{\partial B_x}{\partial z} = N_6,  \label{ms3}
\end{equation}

\begin{equation}
\rho_0 \frac{\partial p}{\partial t} - \gamma p_0 \frac{\partial \rho}{%
\partial t} - \rho_0 (\gamma - 1 ) \left( \frac{\partial Q}{\partial \rho}%
\rho + \frac{\partial Q}{\partial p} p\right) = N_7,  \label{ms4}
\end{equation}

\begin{equation}
\frac{\partial \rho}{\partial t} + \rho_0 \frac{\partial V_z}{\partial z}
=N_8,  \label{ms5}
\end{equation}
where $B_{0x} = B_0\ \sin \alpha$ and $B_{0z} = B_0\ \cos \alpha$; $N_1$, $%
N_2$, $N_4$--$N_8$ are nonlinear terms. The derivatives of the
heating/cooling function $Q$ in equation (\ref{ms4}) are calculated for
unperturbed values of the density and pressure, $\rho=\rho_0$ and $p = p_0$.
According to equation (\ref{omit}), we choose the perturbation $B_z=0$.

Restricting ourselves to keep   quadratic nonlinear terms only, we obtain the
following expressions for the nonlinear right handsides, 
\begin{equation}
N_1 = -\frac{\partial }{\partial z} ( V_z B_x),  \label{N1}
\end{equation}

\begin{equation}
N_2 = - \frac{\partial}{\partial z} \left( V_z B_y \right),  \label{N2}
\end{equation}

\begin{equation}
N_4 = -\rho \frac{\partial V_x}{\partial t} - \rho_0 V_z \frac{\partial V_x%
} {\partial z},  \label{N4}
\end{equation}

\begin{equation}
N_5 = -\rho \frac{\partial V_y}{\partial t} - \rho_0 V_z \frac{\partial V_y%
} {\partial z},  \label{N5}
\end{equation}

\begin{equation}
N_6 = -\rho\frac{\partial V_z}{\partial t} - \rho_0 V_z \frac{\partial V_z} {%
\partial z} - \frac{1}{4\pi} B_x \frac{\partial B_x}{\partial z} - \frac{1}{%
4\pi} B_y \frac{\partial B_y}{\partial z},  \label{N6}
\end{equation}

\begin{equation}
N_7 = -\rho \frac{\partial p}{\partial t} - \rho_0 V_z \frac{\partial p}{%
\partial z} + \gamma p \frac{\partial \rho}{\partial t} + \gamma p_0 V_z 
\frac{\partial \rho}{\partial z},  \label{N7}
\end{equation}

\begin{equation}
N_8 = -\frac{\partial}{\partial z} (\rho V_z).  \label{N8}
\end{equation}

In Eqs. (\ref{N1})---(\ref{N8}), variables $\rho$, $p$, $V_x$ and $B_x$ may
be expressed through the variable $V_z$ from the following relations: 
\begin{equation}
\frac{\partial \rho}{\partial t} = -\rho_0 \frac{\partial V_z}{\partial z},
\label{erho}
\end{equation}

\begin{equation}
\frac{\partial p}{\partial t} = - \rho_0 C_s^2 \frac{\partial V_z}{\partial z%
},  \label{ep}
\end{equation}

\begin{equation}
\frac{\partial^2 V_x}{\partial t^2} = - \cot \alpha \left( \frac{\partial^2}{%
\partial t^2} - C_s^2 \frac{\partial^2}{\partial z^2} \right)\ V_z,
\label{evx}
\end{equation}

\begin{equation}
\left(\frac{\partial^2}{\partial t^2} - C_{A}^2 \cos^2 \alpha \frac{%
\partial^2}{\partial z^2} \right)\ B_x = - B_0 \sin \alpha \frac{\partial^2
V_z}{\partial z \partial t}.  \label{ebx}
\end{equation}

Eqs.~(\ref{a1}) and (\ref{a2}) can be combine to the equation for the
Alfv\'en wave: 
\begin{equation}
{\cal D}_{Az}\ V_y = \frac{1}{\rho_0} \left(\frac{\partial N_5}{\partial t}
+ \frac{B_{0z}}{4\pi} \frac{\partial N_2}{\partial z}\right),  \label{aw}
\end{equation}
where we use the Alfv\'en wave operator, 
\[
{\cal D}_{Az} = \frac{\partial^2}{\partial t^2} - C_{Az}^2\frac{\partial^2}{%
\partial z^2} 
\]
and $C_{Az} = B_{0z}/(4\pi\rho_0)^{1/2}$.

Eqs.~(\ref{ms1})-(\ref{ms5}) can be combined to the equation for
magnetacoustic waves: 
\[
{\cal D}_{Az}{\cal D}_s\ V_z-C_{Ax}^2\frac{\partial ^4V_z}{\partial
z^2\partial t^2}-(\gamma -1) 
\]
\[
\left( \frac{\partial Q}{\partial \rho }+C_s^2\frac{\partial Q}{\partial p}%
\right) {\cal D}_{Az}\int \frac{\partial ^2V_z}{\partial z^2}\ dt= 
\]
\[
=\frac{1}{\rho _0}\left\{ {\cal D}_{Az}\left[ \frac{\partial N_6}{\partial t}%
-\frac{ \partial}{\partial z}\left( C_s^2N_8+\frac 1{\rho _0}N_7\right) %
\right] \right. 
\]
\begin{equation}
\left. -\frac{B_{0x}}{4\pi }\frac{\partial ^2}{\partial z\partial t}\left( 
\frac{\partial N_1}{\partial t}+\frac{B_{0z}}{\rho _0}\frac{\partial N_4}{%
\partial z}\right) \right\} ,  \label{msw}
\end{equation}
where 
\[
{\cal D}_s=\frac{\partial ^2}{\partial t^2}-C_s^2\frac{\partial ^2}{\partial
z^2}, 
\]
$C_s=\sqrt{\gamma p_0/\rho _0}$ is the sound speed and $C_{Ax}=B_{0x}/(4\pi
\rho _0)^{1/2}$.

In the linear limit, the Alfv\'en wave perturbs the variables $B_y$ and $V_y$%
, the magnetosonic waves perturb the variables $B_x$, $V_x$, $V_z$, $\rho$
and $p$. As it follows from (\ref{aw}), the introduction of the
heating/cooling function $Q(p,\rho)$ does not change the propagation of the
Alfv\'en waves, at least till the quadratic nonlinear terms.

Equations (\ref{aw}) and (\ref{msw}) form the governing set of equations for
consideration of quadratically nonlinear MHD wave dynamics in an homogeneous
plasma with nonadiabaticity. Set (\ref{aw}), (\ref{msw}) contains information
about independent dynamics of the MHD waves (Alfv\'en and fast and slow
magnetoacoustic) as well as their nonlinear interaction. This set is an
analogue of the similar set of equations derived in by Nakariakov and
Oraevsky (1995), Nakariakov et al. (1997a,b, 1998) for MHD
waves of finite amplitude in smoothly inhomogeneous cold plasma.

An analysis of the right handsides of equations (\ref{aw}) and (\ref{msw})
shows that if the Alfv\'enic perturbations (i.e. $B_y$ and $V_y$) are
initially absent from the system, they can not be excited by the
magnetoacoustic perturbations, because all terms on the right handside of (%
\ref{aw}) are {\it products} of Alfv\'en and magnetoacoustic variables. This
gives an advantage for an analytical description of the nonlinear dynamics
of the magnetoacoustic waves, allowing to take $V_y$ and $B_y$ zero always
and everywhere. Strictly speaking, this assumption is non-physical, because
Alfv\'enic perturbations are always presented in the realistic plasma as
noise. Consequently, in the natural conditions there is a nonlinear
interaction of Alfv\'en and magnetoacoustic waves. Nevertheless, if the
level of Alfv\'enic noise is initially sufficiently low, the assumption can
be applied to the description of magnetoacoustic waves in some certain
initial stage of the interaction, when nonlinear terms including Alfv\'en
variables are much less than magnetoacoustic terms.

\section{Dispersion relations}

\noindent Supposing $V_z \sim \exp(i\omega t - i k z)$, we obtain the
dispersion relation for the magnetosonic waves from Eq.~(\ref{msw}): 
\[
(\omega^2 - C_{A}^2\cos^2\alpha\ k^2)(\omega^2 - C_s^2 k^2) -
C_{A}^2\sin^2\alpha \omega^2 k^2 
\]
\begin{equation}
+ \frac{i}{\omega} A k^2 (\omega^2 - C_{A}^2\cos^2\alpha\ k^2) = 0,
\label{disp}
\end{equation}
where $C_A = B_0/(4\pi \rho_0)^{1/2}$ and 
\begin{equation}
A = (\gamma-1)\left(\frac{\partial Q}{\partial \rho} + C_s^2 \frac{\partial Q%
}{\partial p}\right).  \label{expA}
\end{equation}

If the heating/cooling function is not taken into account, $A=0$, we have
the well-known dispersion relation for magnetosonic waves, 
\begin{equation}
(\omega^2 - C_A^2 \cos^2\alpha\ k^2)(\omega^2 - C_s^2 k^2) - C_A^2
\sin^2\alpha \ \omega^2 k^2 = 0,
\label{idr}
\end{equation}
which describes dispersion for both fast and slow magnetosonic waves.
Dispersion relation (\ref{idr}) gives that the  phase speed is 
\begin{equation}
C^2_{{\rm fast,\ slow}}  
\frac{1}{2} \left[ C_s^2 +C_A^2 \pm \sqrt{(C_s^2+C_A^2)^2-4 \cos^2 \alpha\
C_s^2 C_A^2}\right],
 \label{psfs}
\end{equation}
where the upper sign corresponds to the fast wave and the lower to the slow
wave. {\ Figure (\ref{fig1}) shows the variation of this speed as a function
of $\alpha$ for different values of plasma beta ($\beta$)}.

Consider the case of the parallel propagation, $\alpha =0$, 
\begin{equation}
(\omega ^2-C_A^2k^2)(\omega ^2-C_s^2k^2+\frac i\omega Ak^2)=0.
\end{equation}
There are fast and slow waves. The thermal instability can be only on the
slow mode, in this case. If $C_A=0$, we obtain the same result as in KT
reference. If $C_s\to 0$ (the coronal case), we have $\omega ^3=-iAk^2$.

Consider transversal propagation, $\alpha = \pi/2$, 
\begin{equation}
\omega^2 \left( \omega^2 - (C_A^2+C_s^2) k^2 + \frac{i A}{\omega} k^2
\right) = 0,
\end{equation}
there is not slow wave now, but the fast wave can be a subject to the
thermal instability.

In the hydrodynamical case, $C_A=0$, the dispersion equation becomes 
\begin{equation}
\omega ^2-C_s^2k^2+\frac i\omega Ak^2=0,  \label{disCA0}
\end{equation}
reproducing the dispersion relation of KT.


\section{Derivation of the evolutionary equation}

Now we take that the influence of the heating/cooling function on
the wave propagation is small, in other words, we suppose that in equation 
(\ref{msw}) the third term on the left handside is much less than
the other linear terms. Also, the quadratically nonlinear terms are taken 
into account. In
the weakly nonlinear, weakly non-adiabatic case, the magnetosonic waves
propagate with the phase speed {\it about} the speeds given by expressions
(\ref{psfs}). The weak nonlinearity and non-adiabaticity lead to {\it slow}
evolution of the waves. (The term \lq\lq slow" means that the typical times
of the nonlinear and non-adiabatic modification are much greater than the
wave period) 

Following one certain magnetoacoustic wave, we change to the running frame of
reference, 
\begin{equation}
\xi =z-Ct,\ \ \ \tau =t,  \label{frame}
\end{equation}
where $C$ is either $C_{{\rm fast}}$ for the fast wave or $C_{{\rm slow}}$
for the slow wave. The variable $\tau $ means the {\it slow} time, which
corresponds to the slow evolution of the wave considered due to the weak
nonadiabaticity and nonlinearity. The dependence of the wave amplitude upon
this slow time $\tau $ represents the difference of the wave propagation in
the presence of the weak nonadiabaticity and nonlinearity from the ideal and
linear case. In this frame of reference, equation (\ref{msw}) is rewritten
as 
\[
-2C(2C^2-C_s^2-C_A^2)\frac{\partial ^4}{\partial \tau \partial \xi ^3}V_z+A%
\frac{C^2-C_A^2\cos ^2\alpha }C\frac{\partial ^3}{\partial \xi ^3}V_z 
\]
\[
=\frac 1{\rho _0}\frac{\partial ^3}{\partial \xi ^3}\left\{ -(C^2-C_A^2\cos
^2\alpha )\left[ CN_6+(C_s^2N_8+\frac 1{\rho _0}N_7)\right] \right. 
\]
\begin{equation}
\left. +\frac{B_0\sin \alpha }{4\pi }C\left( -CN_1+\frac{B_0\cos \alpha }{%
\rho _0}N_4\right) \right\} ,  \label{nmsw}
\end{equation}
where expressions for the nonlinear terms are following: 
\begin{equation}
N_1=-\frac{B_0C\sin \alpha }{C^2-C_A^2\cos ^2\alpha }\frac{\partial V_z^2}{%
\partial \xi },  \label{n1m}
\end{equation}
\begin{equation}
N_4=0,  \label{n4m}
\end{equation}
\begin{equation}
N_6=-\frac{B_0^2}{4\pi }\frac{C^2\sin ^2\alpha }{(C^2-C_A^2\cos ^2\alpha )^2}%
V_z\frac{\partial V_z}{\partial \xi },  \label{n6m}
\end{equation}
\begin{equation}
N_7=-\frac{\gamma (\gamma -1)\rho _0p_0}CV_z\frac{\partial V_z}{\partial \xi 
},  \label{n7m}
\end{equation}
\begin{equation}
N_8=-\frac{2\rho _0}C V_z\frac{\partial V_z}{\partial \xi }.  
\label{n8m}
\end{equation}
Only first order derivatives with respect to the slow time $\tau$ are
kept on the left hand side of equation (\ref{nmsw}) and neglected in the
non-adiabatic and nonlinear terms.
We used expressions (\ref{erho})-(\ref{ebx}) rewritten in the moving frame
of reference (\ref{frame}), 
\[
\rho =\frac{\rho _0}CV_z,\ \ \ p=\rho _0\frac{C_s^2}CV_z,\ \ \ V_x=-\cot
\alpha \frac{C^2-C_s^2}{C^2}V_z, 
\]
\begin{equation}
B_x=\frac{B_0C\sin \alpha }{C^2-C_A^2\cos ^2\alpha }V_z,  \label{eee}
\end{equation}
where only fast dependence on time has been taken into account.

Using (\ref{n1m})---(\ref{n8m}), we can rewrite equation (\ref{nmsw}) as 
\begin{equation}
\frac{\partial ^4}{\partial \tau \partial \xi ^3}V_z+\mu \frac{\partial ^3}{%
\partial \xi ^3}V_z+\varepsilon \frac{\partial ^3}{\partial \xi ^3}V_z\frac{%
\partial V_z}{\partial \xi }=0,  \label{dke}
\end{equation}
where $\mu $ and $\varepsilon $ are the coefficients, 
\begin{equation}
\mu =-A\frac{C^2-C_A^2\cos ^2\alpha }{2C^2(2C^2-C_s^2-C_A^2)},  \label{mu}
\end{equation}
\begin{equation}
\varepsilon =\frac{3C_A^2C^4\sin ^2\alpha +C_s^2(\gamma +1)(C^2-C_A^2\cos
^2\alpha )^2}{2C^2(C^2-C_A^2\cos ^2\alpha )(2C^2-C_s^2-C_A^2)}.
\label{epsilon}
\end{equation}

Integrating equation (\ref{dke}), and assuming that the arbitrary constants
of the integration are zero, we obtain an evolutionary equation describing
the weakly nonlinear dynamics of magnetoacoustic waves in the magnetised
plas\-ma in the presence of the heating/cooling function, 
\begin{equation}
\frac \partial {\partial \tau }V_z+\mu V_z+\varepsilon V_z\frac{\partial V_z%
}{\partial \xi }=0.  \label{ke}
\end{equation}
Equation (\ref{ke}) is an analogue of the Burgers and Korteweg - de Vries
equations (the difference is connected with taking into account of the
effect of nonadiabaticity instead of dissipation or dispersion leading to
appearance of the terms with highest derivatives with respect to the spatial
coordinate), and is a generalisation of the equation obtained in Ref. KT for
acoustic waves.

We note that equation (\ref{ke}) describes the weakly nonlinear and
nonadiabatic evolution of either the fast magnetoacoustic wave or slow
magnetoacoustic wave, but not their nonlinear coupling. The nonlinear
interaction of the magnetoacoustic waves is neglected due to the single wave
approximation applied above. In this approximation, it is supposed that
there initially is only a wave of one certain type in the system, while
another wave is absent. Nonlinear coupling of the MHD waves in the thermal
unstable medium will be considered elsewhere.

Consider dependence of the coefficients $\mu$ and $\epsilon$ of evolutionary
equation (\ref{ke}) on the problem parameters. Actually, there are two
independent parameters in the problem considered, the angle of the wave
propagation $\alpha$ and parameter $\beta$ defined as a ratio of the kinetic
and magnetic pressure in the plasma, $\beta = \frac{2}{\gamma}{C_s^2}/{C_A^2}
$.

{\ Figures (\ref{fig2}) and (\ref{fig3}) show these two coefficients as a
function of $\alpha$ for different values of $\beta$, for fast and slow
waves, respectively. In Figure (\ref{fig2}) can be seen for a given value of
the angle both coefficients increase when $\beta$ decreases but the opposite
occurs for slow waves (see Figure (\ref{fig3})). When $\beta=1.2$ both
coefficients ($\mu C_s^2|A|^{-1}$ and $\epsilon$) are independent of angle $%
\alpha$.}

In the case $B_0=0$ ($C_A=0$, $\beta \to \infty $), we have 
\begin{equation}
C=C_{{\rm fast}}=C_s;\ \ \mu =-\frac A{2C_s^2};\ \ \varepsilon =\frac{\gamma
+1}2.  \label{coefCA0}
\end{equation}
Expressions (\ref{coefCA0}) coincide with the expressions obtained in Ref.
KT. All coefficients are isotropic (independent on the angle of the wave
propagation).

\section{Analysis of wave dynamics}

\noindent Equation (\ref{ke}) is a partial differential equation of the
first order, and can be easily solved analytically in an implicit form (KT), 
\begin{equation}
V_z-F\left( \xi +\frac \varepsilon \mu V_z[1-\exp (\mu \tau )]\right) \exp
(-\mu \tau )=0,  \label{sol}
\end{equation}
where the function $F(x)$ is a profile of the wave in the initial time, $%
V_z(z,0)=F(z)$.

For convenience of the following consideration, we rewrite solution (\ref
{sol}) in the laboratory frame of reference, 
\begin{equation}
V_z - F\left( z - C t + \frac{\varepsilon}{\mu}\ V_z [1-\exp(\mu t)]\right)
\exp(-\mu t) =0.  \label{nsol}
\end{equation}

When the parameter of nonadiabaticity $\mu $ tends to zero, solution (\ref
{nsol}) becomes the well-known simple wave solution (Landau and Lifshitz,
1987) 
\begin{equation}
V_z-F(z-Ct-\varepsilon V_zt)=0  \label{simw}
\end{equation}
describing steepening and breaking of a nonlinear wave in the absence of
dissipation, dispersion and nonadiabaticity. Nonlinearity leads to
generation of second harmonics, carrying out the upward energy shift in the
wave spectrum. In a finite time, there appear infinite gradients in the wave
shape, corresponding to a weak shock wave. If the parameter of nonlinearity $%
\varepsilon $ is also zero (this takes place in the case of a wave with
infinitely small amplitude), solution (\ref{simw}) shows that the wave
propagate without a change of the wave shape. Different signs of the
parameter $\varepsilon $ define where, at the front or rear slope of the
wave (or, more precisely, on the slope with positive or negative sign of
derivative of the function describing the wave profile), the shock formation
does take place.

To obtain the time and position of formation of the shock wave, or the time
and coordinate of where the wave is breaking, it is convenient to use
another implicit form of the solution of equation (\ref{ke}) which can be
easily obtained from solution (\ref{nsol}), 
\begin{equation}
z-Ct+\frac \varepsilon \mu V_z(1-{\rm e}^{\mu t})+f(V_z{\rm e}^{\mu t})=0,
\end{equation}
where the function $f$ is defined implicitly by the initial shape of the
wave, $z=f(V_z)$.

Following Ref. KT, we find that the breaking time is given by the expression 
\begin{equation}
t_{breaking}=-\frac 1\mu \log \left( 1+\frac \mu \varepsilon f^{^{\prime
}}\right) ,  \label{tbreak}
\end{equation}
where $f^{^{\prime }}$ is the value of the derivative of the wave profile at
the inflection point. In the adiabatic case ($\mu =0$), formula (\ref{tbreak}%
) reduces to the well-known formula for a simple wave (Landau and Lifshitz,
1987). In the unmagnetised case $B_0=0$, formula (\ref{tbreak}) reduces to
the formula for acoustic wave breaking time. Since the parameter $%
\varepsilon $ is positive for all angles of propagation for both fast and
slow waves, the shock appears at the slope with $f^{^{\prime }}<0$ for all
angles and both waves. The specific value of the breaking time depends on
the type of the wave considered and the angle of propagation.

Modification of the wave evolution, carried by finite nonadiabaticity,
depends on the sign of the parameter $\mu $. A positive $\mu $ acts
similarly to dissipation, leading to wave decay. In such a case, equation (%
\ref{tbreak}) defines a maximum value of $\mu =\mu _{\max }=\varepsilon
/|f^{\prime }|$ beyond which the damping of the disturbance by
non-isentropic processes occurs in time shorter than the time at which
nonlinear effects go into play, and shocks waves can not be formed. A
negative $\mu $ corresponds to the case of thermal instability, when the
wave amplitude grows in time. (More strictly, in this case, this effect
should be called as a thermal over-stability or oscillatory instability).
According to (\ref{mu}) and Fig.~2a and 3a, the sign of the coefficient of
nonadiabaticity is opposite to the sign of the parameter $A$ introduced in (%
\ref{expA}). Consequently, the amplification of the finite amplitude
magnetoacoustic wave takes place when the parameter $A$ is positive. 


Different regimes of evolution of the initial Gaussian pulse, 
\begin{equation}
V_z(z)=V_0\exp (-z^2/L^2),  \label{gauss}
\end{equation}
where $V_0$ is an amplitude and $L$ is a characteristic length of the pulse
in the initial time $t=0$, are shown on Figure~4. Independently on the sign
of the parameter $\mu $, the leading slope of the pulse is steepening due to
nonlinearity. In the case of the active medium, $\mu <0$, the amplitude of
the pulse is amplified. In the dissipative medium, $\mu >0$, the amplitude
decays. The sketch shown on Fig.~4 is valid qualitatively for both slow and
fast waves. The quantitative differences arise from different angular
dependences of the speed of propagation and nonlinear and nonadiabatic
coefficients (see Fig.~1, 2 and 3).

\section{Astrophysical Applications}

Now, we shall apply the theory developed above to two special astrophysical
plasma systems, a high temperature atomic ($T>10^{4~}K$) plasma and the cold
($T<10^{3}~K$) molecular ISM gas.

\subsection{High Temperature Gas}

An optically thin plasma in the range of temperature $10^{4}<T<10^{7}$ $K$
radiates according to the relation 
\begin{equation}
\Lambda =\rho ^{2}\Lambda _{i}(\frac{T}{T_{i}})^{\eta },\quad
(erg~cm^{-3}~s^{-1})~,  \label{63}
\end{equation}
where the coefficients $\Lambda _{i}$ and the exponent $\eta $ are well
known in the above range of temperature (Vesecky et. al. 1979 ). Therefore,
if one assumes that the plasma heats as $\sim \rho ^{a}T^{b}$ $(erg$ $%
cm^{-3} $ $s^{-1})$ (Dahlburg \& Mariska 1988), then, it becomes
isentropically unstable ($\mu <0$) in the ranges of temperature $1.56\times
10^{4}\leq T<3.16\times 10^{4}$ $K$, $2.51\times 10^{5}\leq T<6.31\times
10^{5}$ $K$ and $2.0\times 10^{6}\leq T<3.16\times 10^{7}$ $K$ for $%
2.51\times 10^{5}\leq T<6.31\times 10^{5}$ $K$ for heating by coronal
current dissipation ($a=b=1$) and heating by Alfv\'{e}n mode/mode conversion
($a=b=7/6),$ and in the range $2.51\times 10^{5}\leq T<6.31\times 10^{5}$ $K$
for a constant heating per unit of mass ($a=1$ and $b=0)$ . Thus, in the
Sun, the enhancement of shock waves formation due to thermal effects will be
relevant in the chromosphere, transition region and corona where the
corresponding e-folding time $t_{e}=2C_{s}^{2}|A|^{-1}$, for thermal
instability is of the order of $5$, $2\times 10^{2}$and $6\times 10^{5}$ $s$%
, respectively. For disturbances with $L/V_{0}\approx 1/|\mu |,$ the
breaking times are of the order of $0.626|\mu |^{-1}$ for a magnetic field
of about $1$ $G$ . This value decreases when the intensity of the magnetic
field increases.

On the other hand, Interstellar Medium regions with temperature in the range 
$2.51\times 10^{5}\leq T<6.31\times 10^{5}$ $K$ (coronal gas) will be
isentropically unstable with e-folding times of the order of $3\times
10^{15} $ $s$ and a breaking time $t_{b}\approx $ $0.609|\mu |^{-1}$ for
disturbances with $L/V_{0}\approx 1/|\mu |,$ if one assumes a constant
heating per unit of mass ($a=1$ and $b=0$) and a magnetic field $B_{G}$ of
about $2\times 10^{-6}G$.

\subsection{Cold Molecular Gas}

For  diffuse and cold ($T<10^{3}$ $K$) ISM molecular gas (Van Dishoeck \&
Black, 1986; Viala, 1986) one may assume, as a first approximation, a
heating by cosmic rays and grain photoelectrons (Iba\~{n}ez and Parravano
1994 and references therein), and a cooling (Hollenbach \& Mc Kee, 1979;
Hollenbach,1988) functions to be of the form

\begin{equation}
\Gamma =H_{0}\rho ~,\quad (erg~cm^{-3}~s^{-1})~,  \label{64}
\end{equation}

\begin{eqnarray}
\Lambda  &=&8.5\times 10^{-5}n^{2}L(CO)+n_{H_{2}}^{2}L(H_{2})+  \nonumber \\
&&n_{H}n_{H_{2}}L(H)~,\quad (erg~cm^{-3}~s^{-1})~,
\end{eqnarray}
where $n_{H}$, $n_{H_{2}}$and $n$ are the number density of the atomic, 
molecular hydrogen, and the total number density of particles, respectively. 
$L(CO)$, $L(H_{2})$ and $L(H)$ are the cooling efficiencies of the $CO$ and $%
H_{2}$ molecules and atomic hydrogen, respectively. In denser molecular
regions \ other heating processes becomes dominant with more complex
dependence of $\Gamma $ on $\rho $ and $T$, detailed study of which will be
reported elsewhere.

The range of temperature (and particle density) \ at which the gas is
isentropically unstable ($A<0)$, is strongly dependent on the ratio $n_{H}/n$
and on the energy input parameter $H_{0}$. In particular, for $n_{H}/n$ $%
=0.9 $ \ these ranges are $115\lesssim T\lesssim 1905$ $K$ ($136\gtrsim n$)
and $90\lesssim T\lesssim 1865$ $K$ ($3\gtrsim n$) for $H_{0}=10^{-2}$ and $%
H_{0}=10^{-4}$, respectively. However, it is found that the above thermal
instability occurs for high values of $n_{H}/n$, i.e. for low concentrations
of $H_{2}$ and therefore, when the transition from atomic to molecular gas
takes place. The range of temperature for which the gas is thermally
unstable decreases when $n_{H_{2}}/n$ increases and is quenched for high
enough values of $n_{H_{2}}/n$ \ ($\gtrsim 0.13$).

The ratios between the breaking and the e-folding time, $t_{b}/t_{e}$ for a
disturbance with Gaussian profile and $L/V_{0}=10^{16}$ $s$, have been
plotted in Figures 5a and 5b, in absence of a magnetic field (dash-dot line)
as well as for the Galactic field $\ $\ value $B_{G}$ $=2\times 10^{-6}$ $G$%
, \ for three different values of the angle $\alpha $, for the fast (a) and
the slow (b) mode . \ \ \ The breaking times show maxima at a temperature of
about $125$ $K$, but the breaking times for the magnetoacoustic modes are
shorter than that corresponding to the hydrodynamic mode. It means \ that
the effect of the magnetic field is to decrease the time when the breaking
of the wave occurs. \ For the fast mode,  the effect of increasing the angle
between the wave number and ${\bf B}$ is to increase the corresponding
breaking time . This dependence  reverses for the slow mode (see Figs. 5a
and 5b).

Figure 5c shows the dependence of $t_{b}/t_{e}$ on $H_{0}$ for the fast
(dash-line ), slow (continuous line) and hydrodynamic (dot-line) mode. The
gross effect of increasing $H_{0}$ is to increase \ the corresponding
breaking times.

The distance between the point where the disturbance is originated to the
point where the shock is formed, is of the order of $C_{s}t_{b}$. Therefore,
from above results one may conclude \ that strong inhomogeneities are
expected to be formed at time scales and scale-lengths \ shorter than those
for which the molecular gas becomes thermally unstable.

\section{Discussion}

We considered dynamics of magnetohydrodynamic waves in a nonadiabatic plasma
taking into account effects of finite wave amplitude. The nonadiabaticity
was modelled by a general cooling/heating function of thermodynamic
parameters of the plasma. Depending on a general structure of the
nonadiabatic function, there are two different regimes of the wave evolution
in the system, either amplification or decay. The nonlinearity, through
generation of highest harmonics, leads to the wave steepening and,
therefore, to formation of shock waves. Our analysis was restricted by this
shock formation time. In the linearly unstable case (usually called the
thermal instability), the formation of the shock takes place quicker than in
adiabatic case. Consequently, nonlinearity depresses the thermal instability
of magnetoacoustic waves, whose growth is accompanied by the wave steepening
and lasts until the shock has been formed.

In the single wave approximation, we have derived an evolutionary equation
describing development of weakly nonlinear magnetoacoustic waves in a weakly
nonadiabatic medium. This equation is valid for one magnetoacoustic wave
(either slow or fast) providing that another wave is absent from the system.
Of course, in realistic situation, both slow and fast waves are excited
simultaneously. Nevertheless, the difference in their propagation speeds
gives us a possibility to restrict our consideration by the single wave
approximation, because the waves occupy different locations in time after
the excitation (they are spatially de-tuned from each other). Since the
magnetic field brings spatial anisotropy in the system considered,
parameters of the magnetoacoustic wave propagation (the speed, the rate of
growth or decay, the breaking time) depend upon the direction of the wave
vector. The third magnetoacoustic mode, the Alfv\'en wave, is not subject to
the influence of the nonadiabaticity (as far as the quadratic nonlinear
terms approximation holds). This does not exclude a possibility of the
indirect effect of the nonadiabaticity of the plasma on the Alfv\'en wave
through nonlinear coupling with the magnetoacoustic wave amplified by
thermal instability. The nonlinear excitation of the Alfv\'en waves is
described by Eq.~(\ref{aw}). According to above discussion, the active
non-adiabaticity of the plasma leads to excitation of magnetoacoustic waves
perturbing the variables $B_x$, $V_x$, $V_z$, $\rho $ and $p$. Expressing
all variables through $V_z$ by (\ref{erho})-(\ref{ebx}), we can re-write the
nonlinear Alfv\'en wave equation (\ref{aw}) for two variables, $V_y$ and $%
V_z $: 
\[
\left( \frac{\partial ^2}{\partial t^2}-C_{Az}^2\frac{\partial ^2}{\partial
z^2}\right) V_y= 
\]
\begin{equation}
\frac{\partial V_z}{\partial z}\frac{\partial V_y}{\partial t}-\frac
\partial {\partial t}\left( V_z\frac{\partial V_y}{\partial z}\right) -C_A^2%
\frac{\partial ^2}{\partial z^2}\left( V_zD_t^{-1}\frac{\partial V_y}{%
\partial z}\right) ,  \label{awgen}
\end{equation}
where $D_t^{-1}$ is an inverse differential operator. In the case of a
harmonical magnetoacoustic wave, Eq.~(\ref{awgen}) describes generation of
the Alfv\'en wave by a parametric instability. But, in the case considered,
the function $V_z$ is prescribed by expression (\ref{sol}). The spectrum of
the exciting magnetoacoustic wave is much reacher than harmonical and
evolves in time and space. However, qualitatively, we can expect that
Alfv\'en waves are nonlinearly generated by the thermal instability through
the discussed mechanism. This phenomenon has been omitted from this
consideration.

The theory developed here may be applied, in particular, to an
interpretation of origin of density, temperature and magnetic field
variations in astrophysical plasmas, which can be associated with the
magnetoacoustic weak shock waves (see, Krasnobaev, 1975). With respect to
the previously analysed case of the unmagnetised plasma, the presence of the
magnetic field gives several advantages for the observational registration
of these phenomena. Firstly, in the vicinity of the shock front, the sharp
inhomogeneity of the absolute value of the magnetic field takes place, which
can be directly registered by analysis of variation of parameters of
gyro-emission and Zeeman splitting. Even if the wave amplitude is
sufficiently small, the sharp gradients in the absolute value of the
magnetic field in the vicinity of the shock are accompanied by high
densities of the electric currents (the transversal component of the current
density is proportional to the spatial derivative of the magnetic field
perturbation, $j_y=-(c/4\pi )\partial B_x/\partial z$). These current
concentrations (or current sheets, in the ideal limit) can be regions where
particle acceleration takes place. Secondly, slow and fast magnetoacoustic
perturbations, initially excited somehow, propagate with different speeds
and growth rates and, therefore, form different spatial structures at
different distance from their source. According to formula (\ref{eee}),
perturbations of the magnetic field are either positively or negatively
correlated with perturbations of the plasma density, depending on the type
of the mode (slow or fast), which gives us a possibility to distinguish fast
and slow modes in analysed data.

From the point of view of the nonlinear wave dynamics, propagation of the
MHD waves in the optically thin plasma is a very attractive problem, because
the plasma is acting in this case as an active medium, supplying energy to
the waves. The simplest model considered here can be further developed. For
example, high frequency dissipation (modelled by introduction of a Burgers
term in equation (\ref{ke})), will compete with the amplification and
depress nonlinear generation of highest harmonics. When these phenomena are
in balance, there exist dissipative stationary waves with ``saw-teeth''
shape. Also, the generation of the second harmonics and so the energy 
transfer to small scales can be depressed by nonlinear dispersion (Whitham,
1974). This process is based upon  either cubic nonlinearity, or quadratic nonlinearity
through  cascade self-interaction.
The last mechanism has two stages: nonlinear generation of second harmonics
and interaction of the second harmonics with a backward wave. In the case
of magnetoacoustic waves, both mechanisms are acting together (Nakariakov et al., 1997b)
leading to the appearance of a cubically nonlinear term in the evolutionary
equation. Obviously, the cubically nonlinear effects should be taken into
account only when quadratically nonlinear processes are suppressed due to,
e.g., linear high frequency dispersion.
Effects of dispersion (connected possibly with electron inertia or,
which is more important for astrophysical applications, with inhomogeneity
of the plasma) can be modelled by introduction of a Korteweg - de Vries (or
Benjamin - Ono or Leibovich - Roberts, depending on the specific situation)
term in evolutionary equation (\ref{ke}). Particularly, the dispersion of
the medium can give a possibility for existence of autosolitons and
stochastic dynamical regimes (see, e.g. Engelbrecht, 1989, Nakariakov and
Roberts 1999). These phenomena can be registered in astrophysical plasmas and
be applied to interpretation of the observations.

Finally, we must remark that in atomic as well as in molecular plasma
occurring in astrophysics, many other physical effects likely go into play,
such as inhomogeneity in the background due to self-gravitation and/or to
radiation transport, and dissipative effects ( viscosity, thermal
conduction, ionization-recombinations and chemical reactions). These phenomena
have to\ be taken into account in more developed models. However, the above
results indicate that inside thermally unstable plasma regions,
substructures can be originated by nonlinear effects in times shorter than
an e-folding time for evolving the isentropic thermal instability.

\section{Acknowledgements}

VMN is grateful to Prof Bernard Roberts and Dr Mikhail Ruderman for a number of
stimulating discussions. CAMB and MHIS thank to the CDCHT of Universidad de
los Andes for its support through projects C-931-99-05-B and C-912-98-B-05, respectively.

\clearpage

\figcaption[fig1.eps]{The linear speeds $C/C_s$ as a function of $\alpha $ for
different values of $\beta $. Continuous and dashed lines correspond to the
fast and slow modes, respectively. \label{fig1}}

\figcaption[fig2a.eps,fig2b.eps]{(a) The non-adiabaticity ($\mu
C_s^2|A|^{-1}$) and (b) the nonlinearity $\epsilon $ parameters as functions
of $\alpha $ for different values of $\beta $ \ corresponding to the fast
wave. \label{fig2}}

\figcaption[fig3a.eps,fig3b.eps]{As Figure~2 for the slow wave. \label{fig3}}

\figcaption[fig4a.eps,fig4b.eps]{Development of the shape of the initial Gaussian
magnetoacoustic pulse with the width $L$ situated initially $t=0$ in $z=0$,
and the indicated time in units of $L/V_0$. In the active medium (a) ($\mu <0
$) the pulse is amplified. In the dissipative medium (b) $\mu >0$ \ the
pulse decays. \label{fig4}}

\figcaption[fig5a.eps,fig5b.eps,fig5c.eps]{The breaking times for the fast\ (a) and slow mode (b) as
a function of temperature $T$, for  $H_{0}=10^{-3}$, $B=0$
(dash-dot line), $B=B_{G}$, $n_{H}/n=0.9$ and  three values of
the angle  $\alpha =\pi /36$ (dash line ), $\pi /4$ (solid line) and  $17\pi /36$ ( dot line). Fig.5c
is as Fig. 5a for two different values of $H_{0}$, $10^{-3}$ and $10^{-4}$.
\label{fig5}}

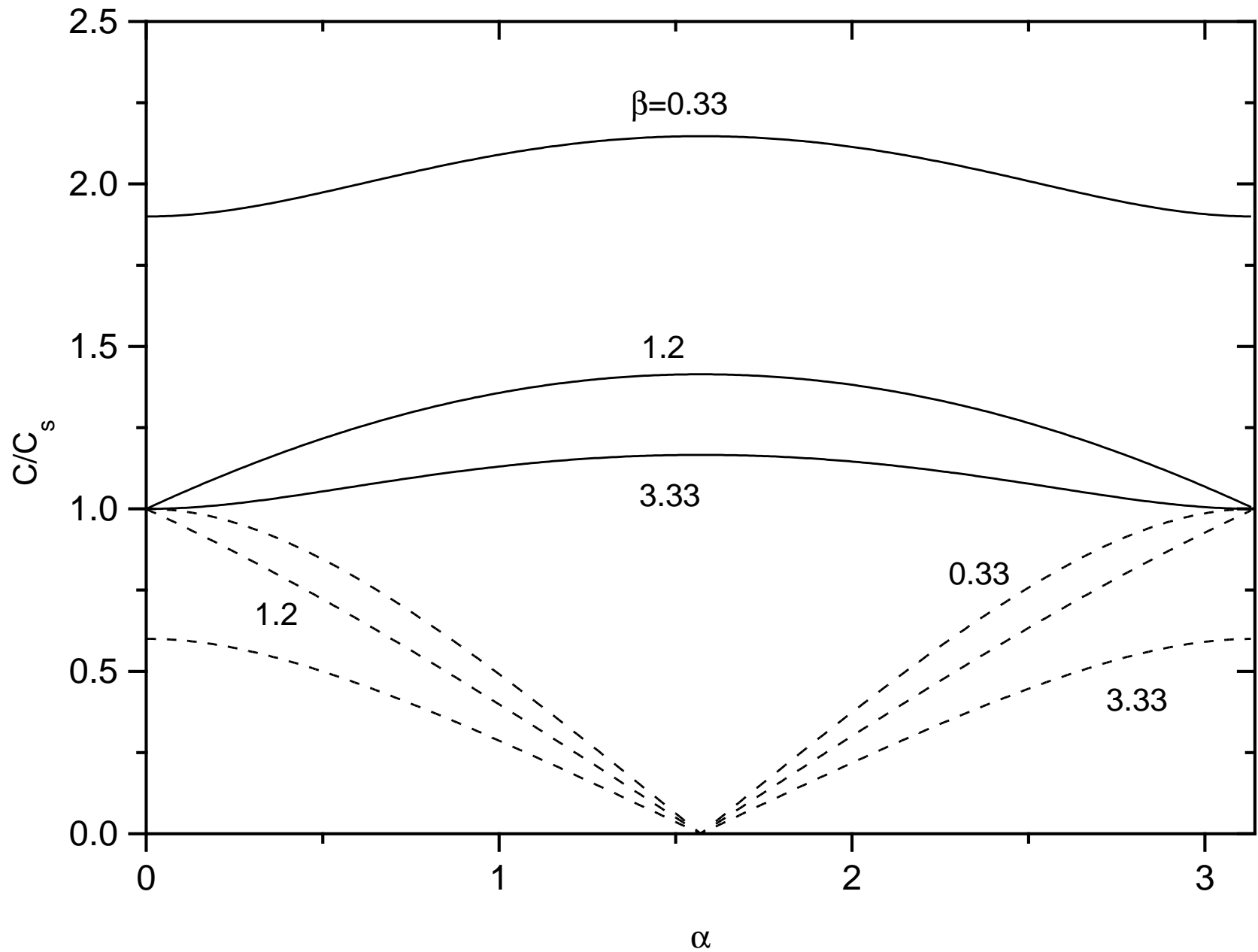
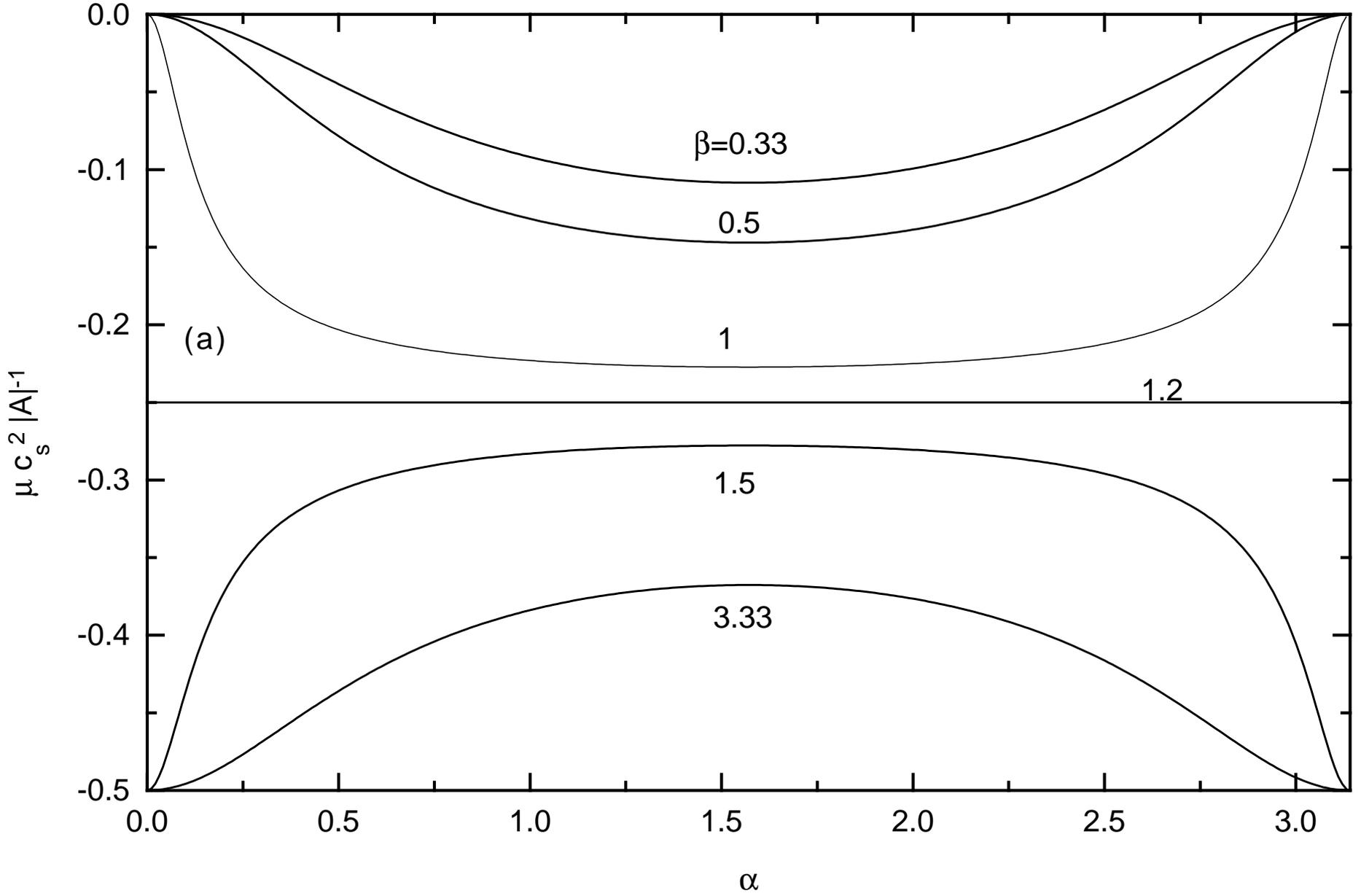
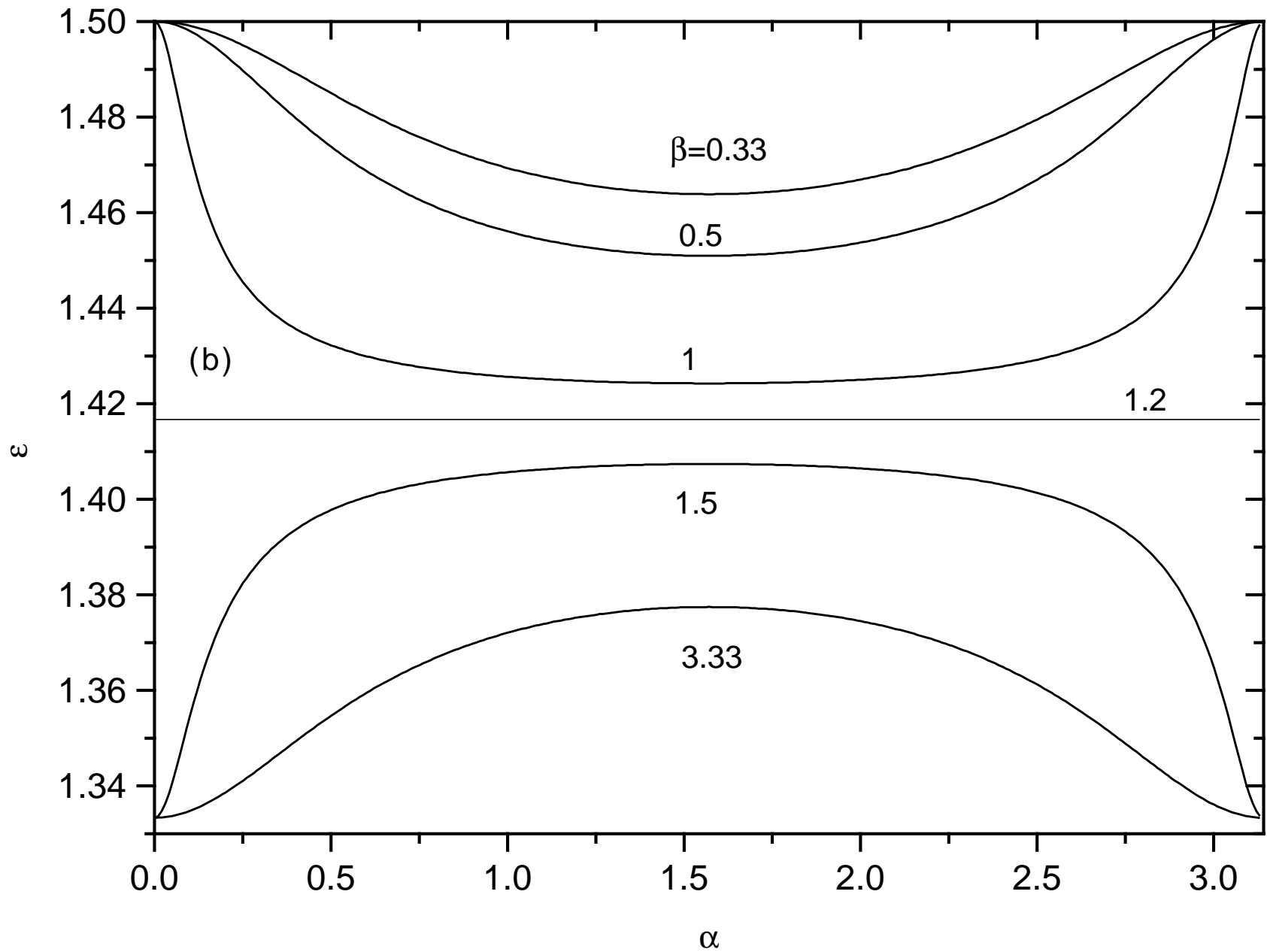
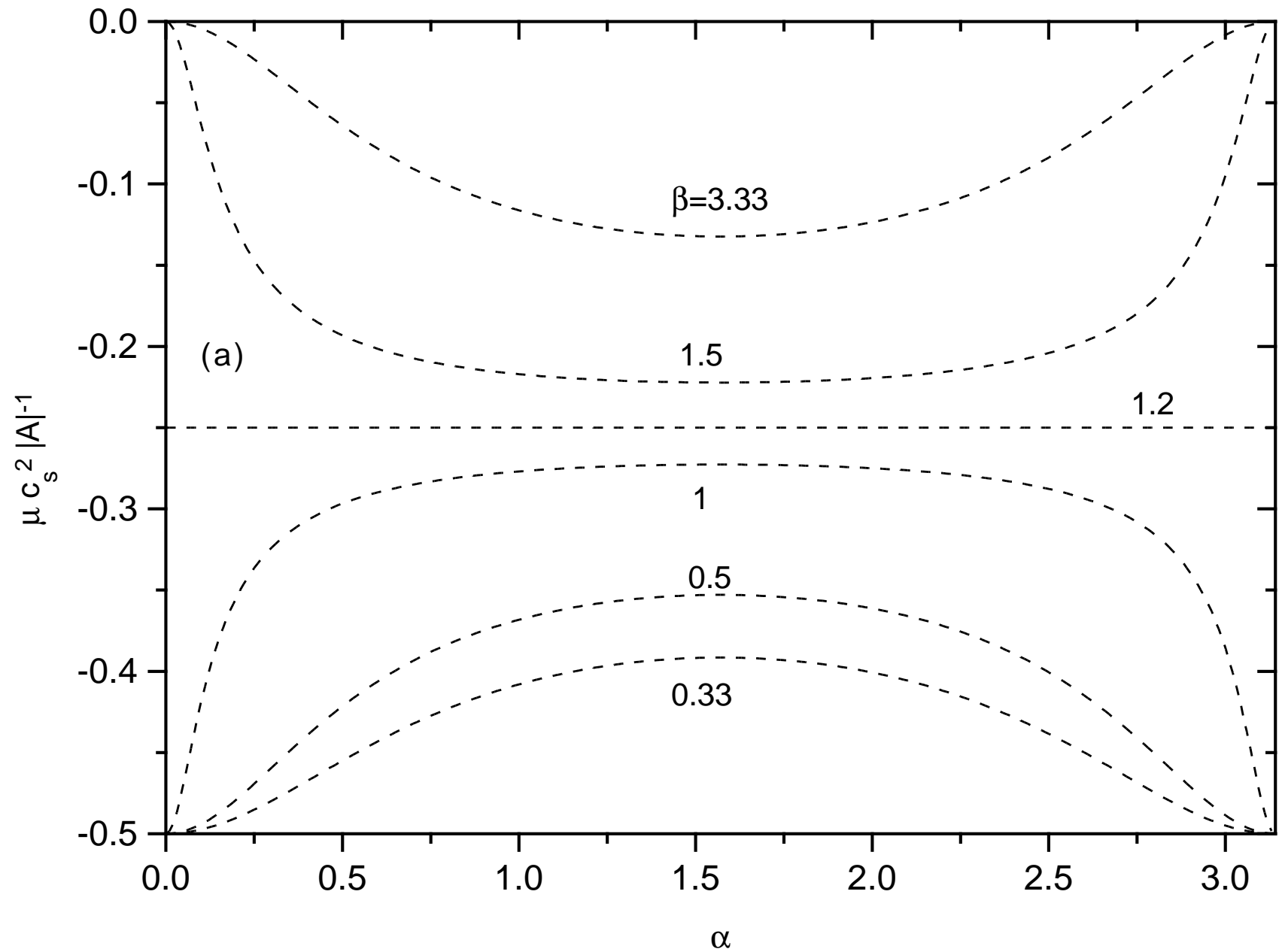
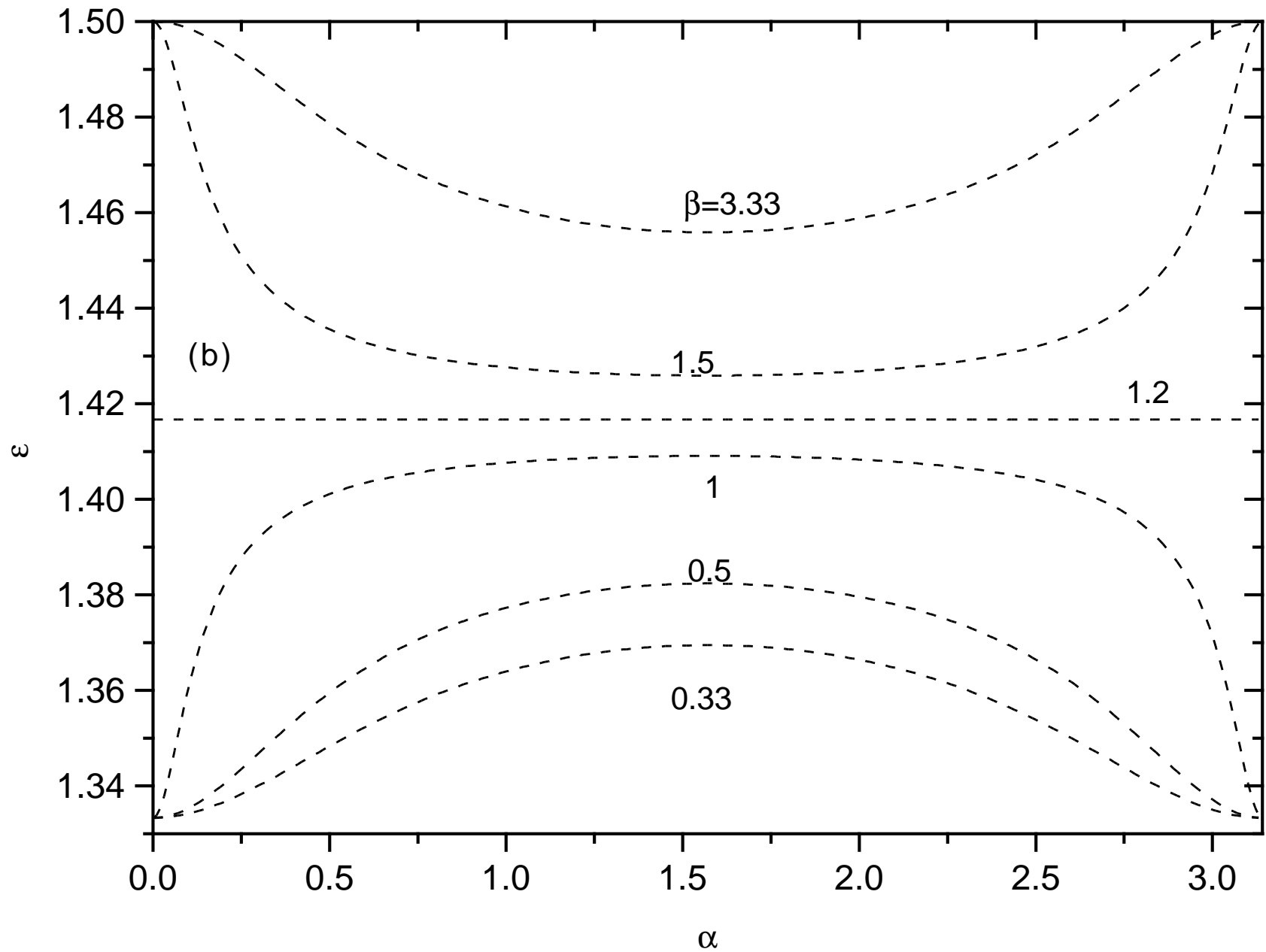
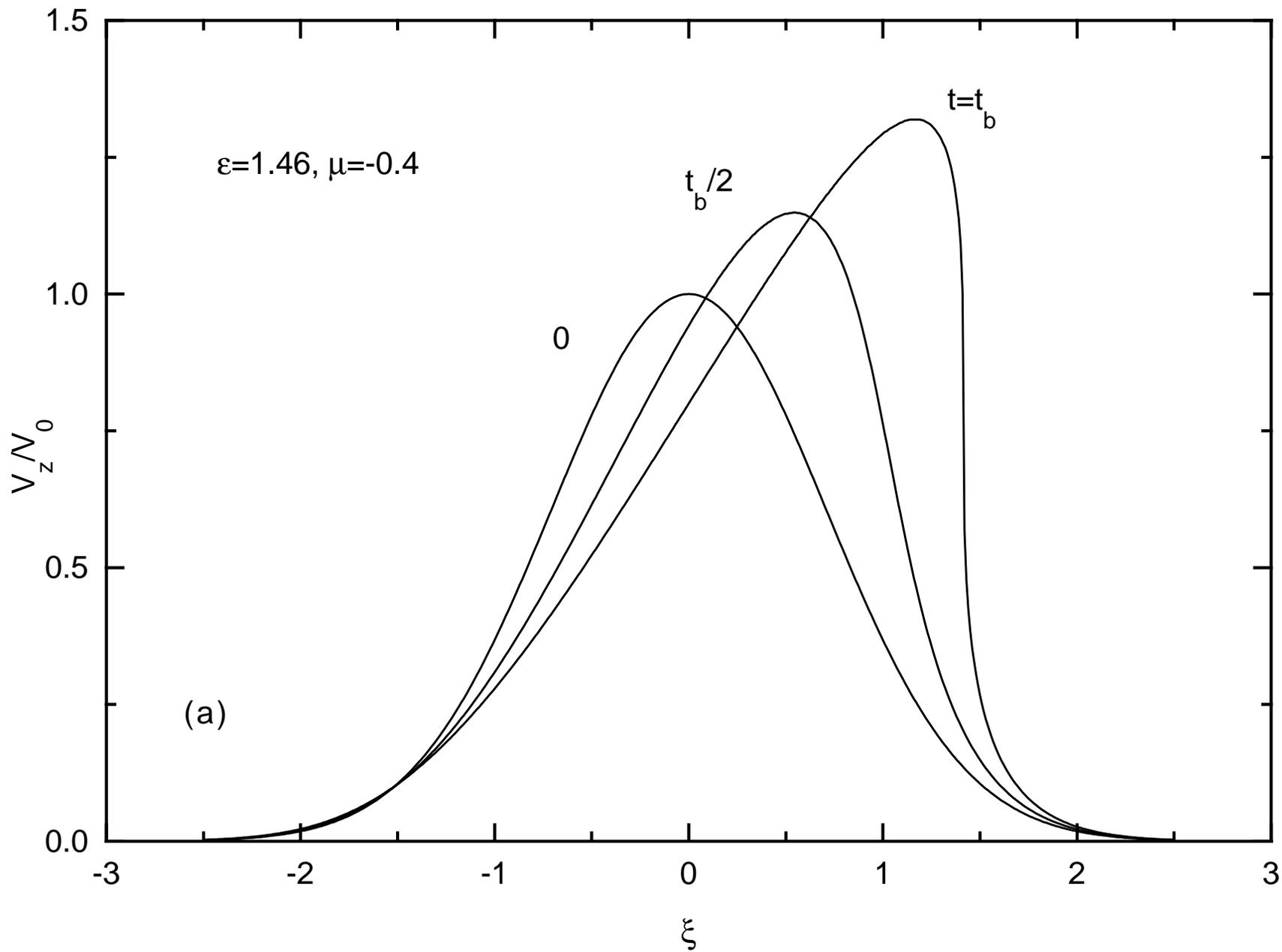
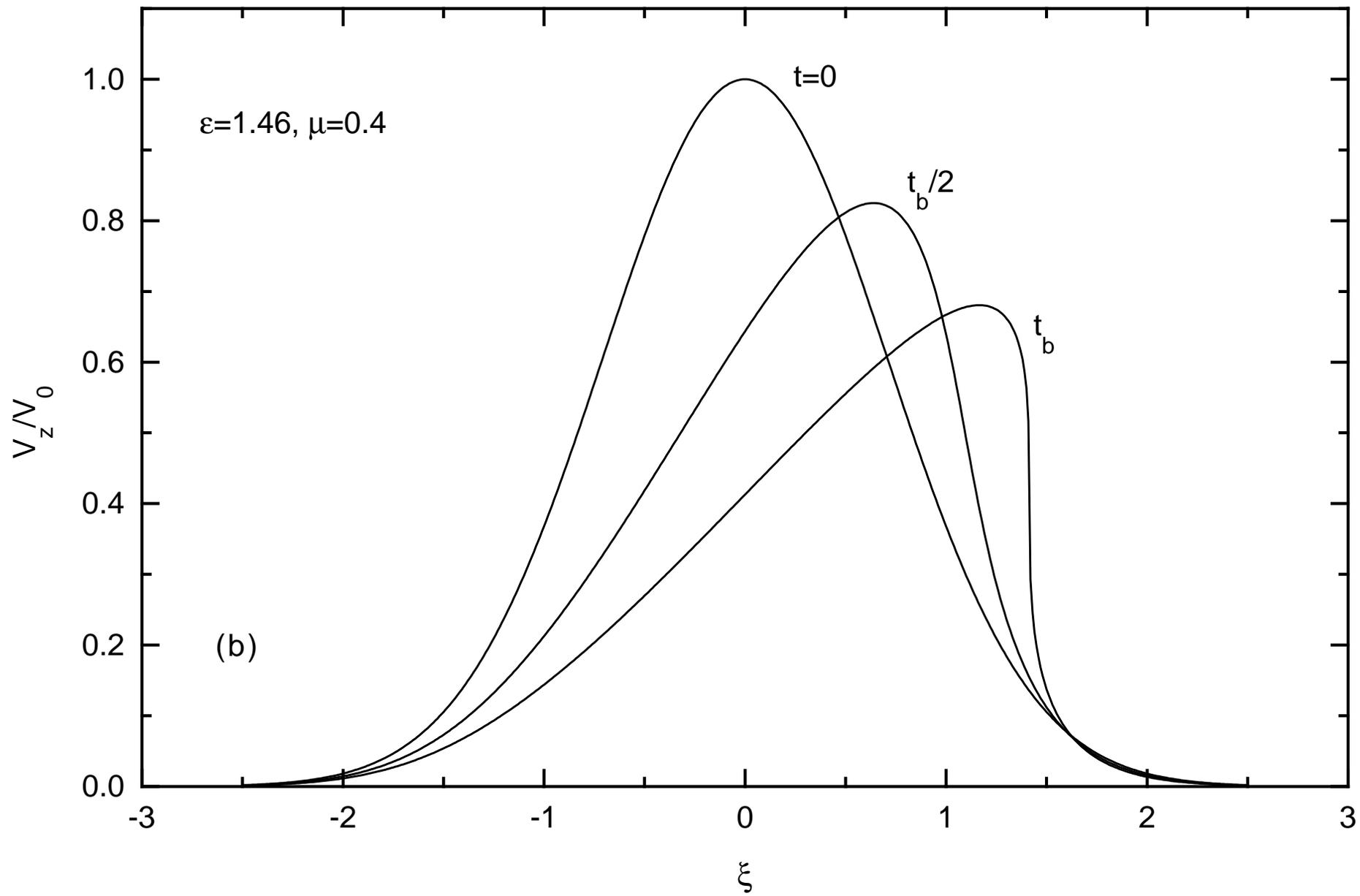
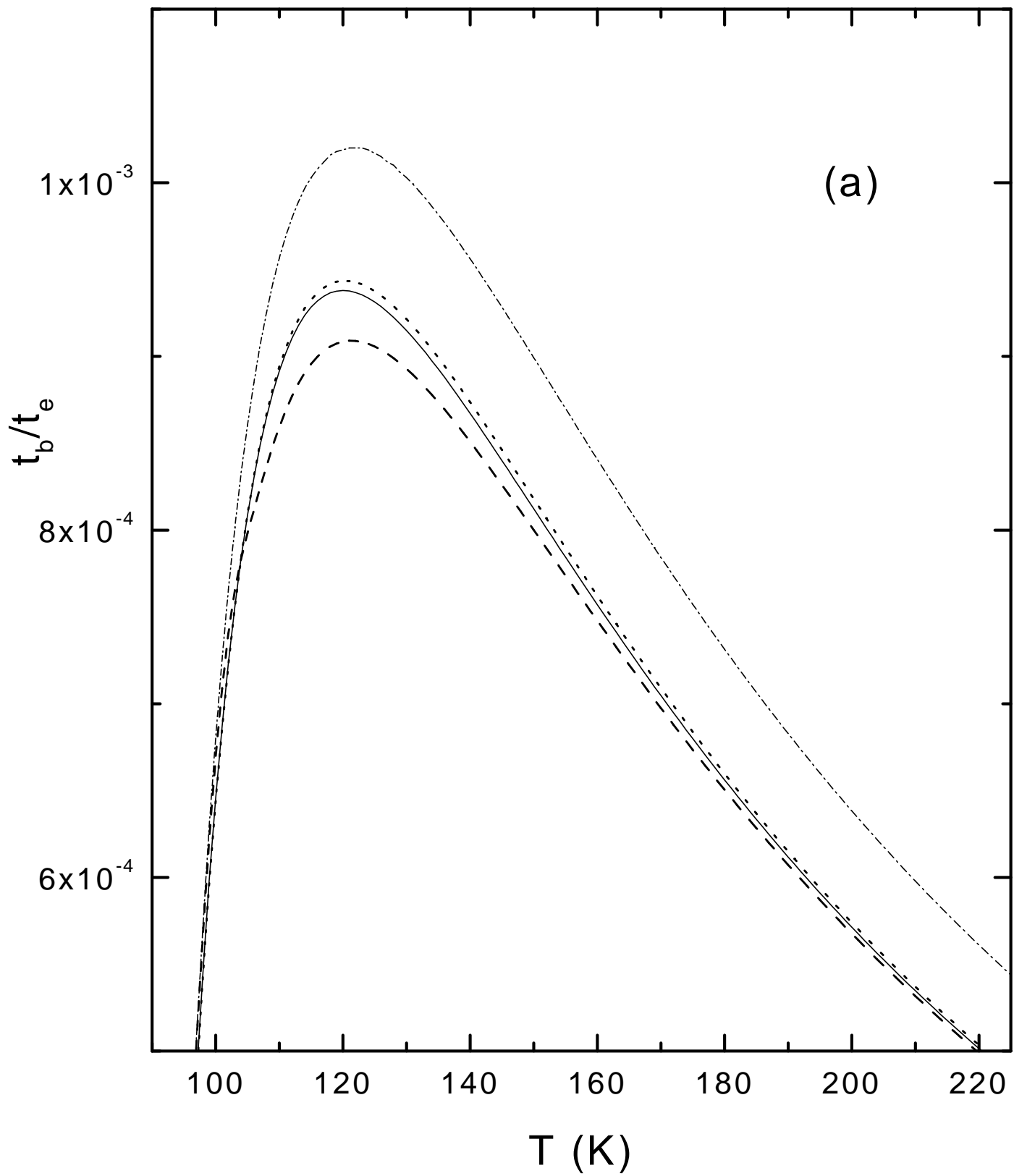
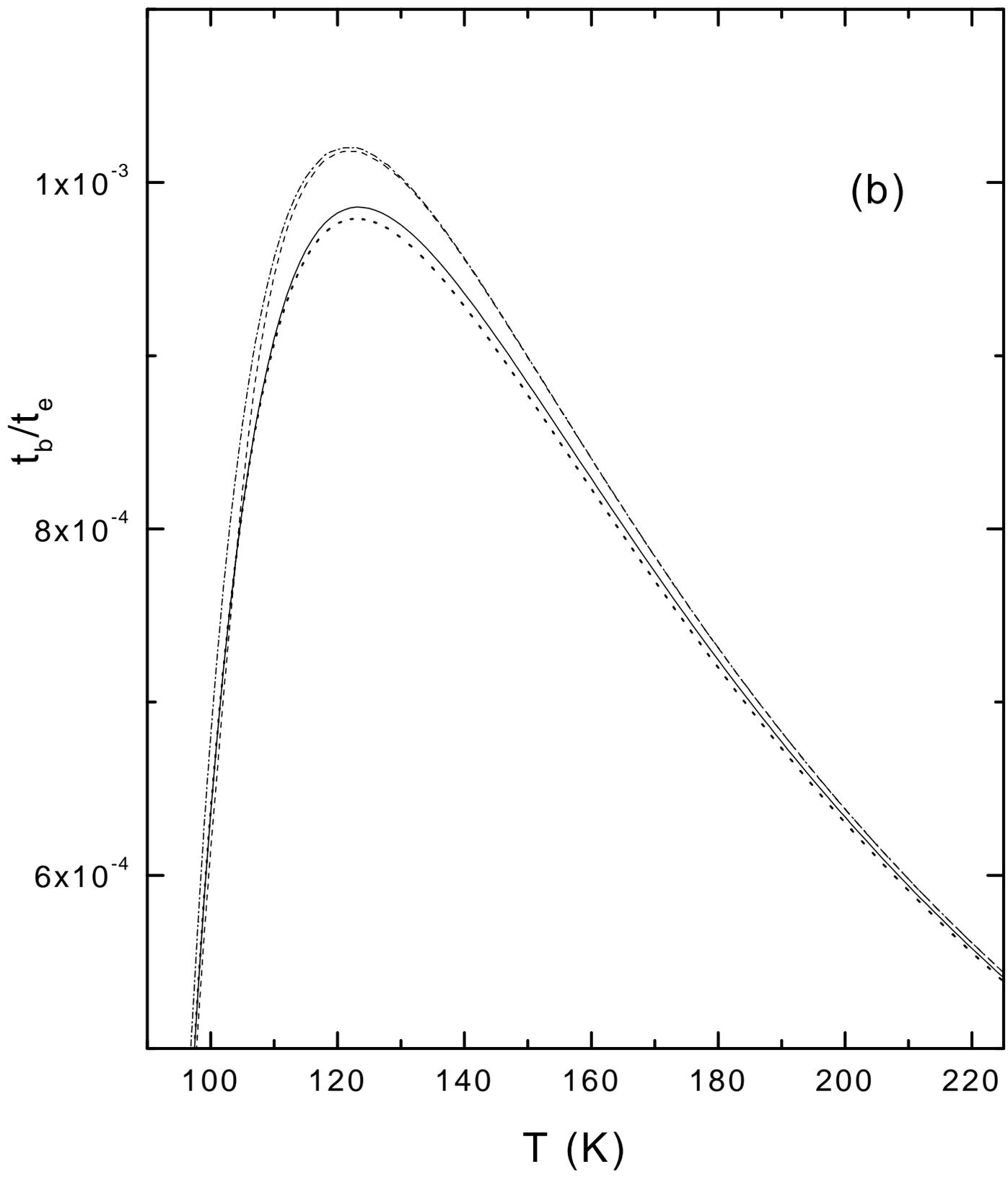
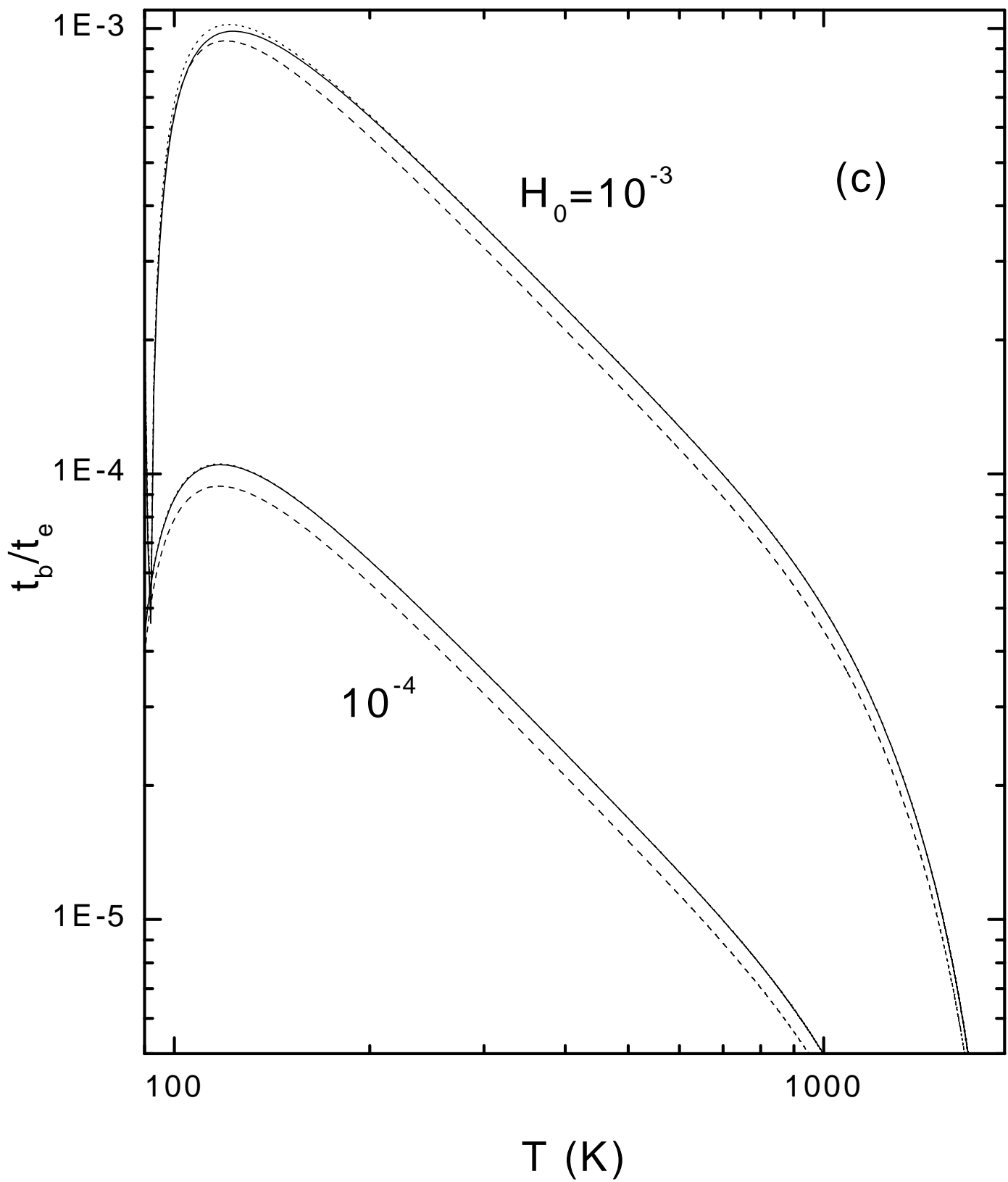
\end{document}